\documentclass[aps,prl,twocolumn,groupedaddress]{revtex4}
\usepackage{amsmath}
\usepackage{graphicx}
\usepackage{subfigure}
\usepackage{epstopdf}
\usepackage{epsfig}
\usepackage{amssymb}
\usepackage{bm}
\usepackage{bbm}
\newcommand{\beq}{\begin{equation}}
\newcommand{\eeq}{\end{equation}}
\newcommand{\ov}{\overline}

\newcommand{\al}{\alpha}

\usepackage{color}

\begin{document}

\title{ Top forward-backward asymmetry and the CDF $\bm{Wjj}$ excess 
\\ in leptophobic  $\bm{U(1)^{'}}$ flavor  models }

\author{P. Ko}
\affiliation{School of Physics, KIAS, Seoul 130-722, Korea}

\author{Yuji Omura}
\affiliation{School of Physics, KIAS, Seoul 130-722, Korea}

\author{Chaehyun Yu}
\affiliation{School of Physics, KIAS, Seoul 130-722, Korea}


\begin{abstract}
\noindent
We construct anomaly-free leptophobic $U(1)^{'}$ flavor models 
with light $Z^{'} (\sim 145$ GeV). 
In order to allow renormalizable Yukawa interactions  
for the standard model chiral fermions,  
new Higgs doublets with nonzero $U(1)^{'}$ charges are introduced.  
Then the neutral (pseudo)scalar Higgs bosons as well as $Z^{'}$ 
contribute to the $t\bar{t}$ and the same sign top pair productions 
[$\sigma(t\bar{t})$ and $\sigma(tt)$], 
and one can evade the strong constraint from $\sigma(tt)$. 
The top forward-backward asymmetry ($A_\textrm{FB}$) and 
$Wjj$ excess at CDF could be accommodated by 
$A_\textrm{FB}^\textrm{New} = 0.084 \sim 0.12$ and 
$\sigma(W j j) \lesssim O(10)$ pb $\times \sin^2 2 \beta$.
\end{abstract}

\pacs{}

\maketitle


The top forward-backward (FB) asymmetry at the Tevatron ($\equiv A_\textrm{FB}^t$) 
has been an interesting issue in particle physics recently, since this quantity has deviated from the standard model (SM) prediction ($0.051\pm0.06$)  
\cite{Antunano:2007da} at the level of $\sim 2 \sigma$ for the last few years.  
There have been considerable attempts to explain the top FB asymmetry 
in some extensions of the SM.
At present, it is premature to tell which model is favored over other models.
However, one common property of the proposed solutions is that one has to  
introduce nontrivial flavor structures in new physics in order to distinguish 
top quarks from light quarks.  
For example, the model by Jung et al. uses light $Z^{'}$ ($\sim$ 160 GeV) 
to explain the top FB asymmetry with larger flavor changing couplings than
flavor conserving couplings of $Z^{'}$ to the right-handed up type quarks
\cite{Jung:2009jz}.
Recently, this model has been excluded by the CMS study of the same sign 
top pair production \cite{Chatrchyan:2011dk}.  However, in this article, 
we will show that there are additional contributions from (pseudo)scalar 
Higgs bosons which must be included in complete models of such light 
$Z^{'}$ with flavor-dependent couplings, which is one of our main themes.

Another interesting observation that might be related with light $Z^{'}$ is
the CDF $Wjj$ excess \cite{Aaltonen:2011mk},
one possible interpretation of which is $p\bar{p} \rightarrow WZ^{'}$
followed by $Z^{'} \rightarrow jj$ with $\sigma (WZ^{'} ) \sim 4$ pb
and $m_{Z^{'}} \sim 140$ GeV \cite{Cheung:2011zt,Ko:2011ns}
(see Ref.~\cite{Plehn:2011nx} for an alternative interpretation 
within the SM). 
This excess, however, was not confirmed by the D0 Collaboration
\cite{Abazov:2011af}, and more investigation is necessary for understanding
this discrepancy. Further data from LHC for this channel will also shed
light on this issue.

Such a relatively light gauge boson with rather strong couplings 
to the SM quarks can evade all the strong constraints from colliders 
only if $Z^{'}$ couples dominantly to the SM quarks, and so $Z^{'}$ better be 
leptophobic.  In the models of Ref.s~\cite{Jung:2009jz}, the top FB asymmetry 
requires a large coupling of $u_R - t_R - Z^{'}$, 
and so one has to consider flavor dependent $U(1)^{'}$ couplings.
Since the massive spin-1 particle should be a gauge boson associated with 
some local $U(1)^{'}$ gauge symmetry, one has to identify the charge 
assignments of the SM fermions under this new $U(1)^{'}$. 
Most likely such a flavor dependent leptophobic $U(1)^{'}$ will be anomalous, 
and one has to introduce additional fermions in order to cancel 
all the gauge anomalies.  If the SM quarks carry the extra $U(1)^{'}$ charges, 
one has to introduce additional $SU(2)_L$ doublet Higgs fields with nonzero 
$U(1)^{'}$ charges in order to write down the renormalizable Yukawa couplings 
for the up quarks. Such new Higgs fields cannot be arbitrarily heavy, and 
they can affect generically the top FB asymmetry 
by the $t$-channel exchanges of (pseudo) neutral/charged 
Higgs bosons, which is one of the salient features of this work.

This article is organized as follows. We first describe the flavor dependent 
leptophobic $U(1)^{'}$ model and introduce new $U(1)^{'}$-charged Higgs fields 
for realistic renormalizable Yukawa couplings for the SM quarks, and 
new fermions for the anomaly cancellation, respectively. 
Then the couplings of the SM fermions to the new $Z^{'}$ and 
the (pseudo) neutral/charged Higgs fields are derived and are used for 
studying the top FB asymmetry at the Tevatron, the same sign top pair 
production at the LHC and the CDF $Wjj$ excess.  We will find that 
the interference between the $Z^{'}$ and the neutral Higgs bosons generally 
improves the overall description of the $t\bar{t}$ production cross section,  
the top FB asymmetry and their $M_{t\bar{t}}$ distributions at the Tevatron, 
while it reduces the production cross section for the same sign top pair at the 
LHC and makes the light $Z^{'}$ still viable.  
Note that the neutral Higgs contribution should be included in the complete 
models with $Z^{'}$, and it is not consistent to include only the $Z^{'}$ 
contributions to those observables. 
 

Since there are very stringent constraints on the new sources of 
flavor-changing neutral-current from  
$K^0$-$\ov{K^0}$, $B_{d(s)}^0$-$\ov{B_{d(s)}^0}$ and $D^0$-$\ov{D^0}$ mixings, 
it would be simplest to assume that only the right-handed (RH) up quarks 
($U_R^i$ with $i=1,2,3$ being the generation index) carry flavor dependent 
$U(1)^{'}$ charges,  whereas the left-handed up and down quarks 
($U_L^i$ and $D_L^i$ ) and the RH down quarks ($D_R^i$) are either neutral 
or universally charged under $U(1)^{'}$.    
Let us define charges of the RH down quarks in the interaction eigenstate basis as $u_i$.
Note that this interaction eigenstates will differ from the mass eigenstates in general,
so that such flavor-dependent $U(1)'$ charge assignment will 
give flavor-dependent couplings with $U(1)'$ gauge boson 
after the basis rotation into the mass eigenstates $\Hat{U}_R^i$:  
\begin{equation}
{\cal L} \supset g'Z'^{\mu} \left \{ (g^u_R)_{ij} 
\ov{\Hat{U}_R^i} \gamma_{\mu} \Hat{U}_R^j \right \}.
\end{equation}
Here $(g_R^u)_{ij}$ are defined as $(R^{u})_{ik}  u_k (R^{u})^{\dagger}_{kj}$, 
where $(R^{u})_{ij}$ is the $3\times 3$ unitary matrix rotating 
the RH up quark fields in order to diagonalize the up quark mass matrix: 
$M^u_{\rm diag} = L^u M^u R^{u\dagger}$. 

$(g^u_R)_{ij}$ could generally have nonzero off-diagonal elements. 
In particular, nonzero $(g^u_R)_{13}$ for the top FB asymmetry 
could be large, whereas $(g^u_R)_{12}$ that would contribute  to the 
$D^0$-$\ov{D^0}$ mixing  is suppressed, by adjusting the Yukawa couplings 
and the vacuum expectation values 
of Higgs doublets depending on $U(1)^{'}$ charge assignments.
Furthermore there reside CP violating phases in $(g_R^u)_{ij}$ whose effects 
would be visible in the same sign top pair production through interference 
among $Z^{'}, h$ and $a$ contributions in the $t$-channel. 

Besides, in order to get renormalizable Yukawa couplings for up-type quarks, 
we have to introduce  new $U(1)^{'}$-charged Higgs doublets, whose $U(1)^{'}$
charges will depend on the $U(1)^{'}$ charge assignments of the SM fermions.  
For example, the assignment satisfying  $u_1=u_2=0$ and $u_3 \neq 0$ 
requires at least one extra Higgs ($\equiv H_3$) whose $U(1)'$ charge is $u_3$. 
In a generic case, $u_1 \neq u_2$ and $u_2 \neq u_3$, four extra Higgs
are required, and if one of $u_i$ is zero, three extra Higgs doublets are 
necessary for renormalizable Yukawa interactions for the up-type quarks.   
Some of the (pseudo) neutral and charged Higgs bosons will have masses 
around a weak scale, and they also contribute to the top FB asymmetry and 
flavor changing processes through Yukawa couplings. 
This is because Yukawa couplings of Higgs fields with quarks could be 
flavor-changing couplings.  

As an example, let us consider one simple case, 
\begin{equation}
(u_1,~u_2,~u_3)=(0,~0,~1)
\label{assign}%
\end{equation}
which corresponds to the two-Higgs doublet model: one is SM Higgs, $H$, 
to couple not only with $U_R^1$ and $U_R^2$, but also down-type 
quarks and leptons, and the other is $H_3$ with $U(1)'$ charge, $+1$, 
and couples with $U_R^3$.  Let us define their vacuum expectation values as 
$(\langle H \rangle,~\langle H_3 \rangle) = 
(v \cos \beta / \sqrt{2}, ~v \sin \beta / \sqrt{2})$. 

In our model, the lightest pseudo-scalar Higgs ($a$) and 
the lightest charged Higgs ($h^{\pm}$) will play important 
roles for the top FB asymmetry and the CDF $Wjj$ excess.
Their masses are derived from the following interaction:   
\begin{equation}
\mu H_3^{\dagger} H \Phi^{(1/q_{\Phi}}  +h.c.,
\end{equation}
where $\Phi$ is a SM-gauge singlet with $U(1)'$ charge, $q_{\Phi}$, 
that breaks $U(1)'$ spontaneously.
In order to realize large pseudo-scalar and charged Higgs masses, 
we have to define $q_{\Phi}=1$ or $1/2$, and the mass dimension of 
$\mu$ will be either 1 or 2 accordingly.

Then, in the mass basis,  the Yukawa couplings of lightest neutral 
scalar Higgs ($h$), charged Higgs $h^\pm$ and pseudoscalar $a$ with 
$\Hat{U}_R^i$ and $\Hat{U}_L^j$ are described by $g_R^u$ and $m_i^u$: 
\begin{widetext}
\begin{equation}
{\cal L} = - Y^u_{ij}  \ov{ \Hat{U}_{Li}} \Hat{U}_{Rj} h 
+ Y^{u-}_{ij} \ov{\Hat{D}_{Li}}h^- \Hat{U}_{Rj}
+ i a Y^a_{ij} \ov{ \Hat{U}_{Li}} \Hat{U}_{Rj} + h.c.,
\end{equation}
\end{widetext}
where the Yukawa couplings are given by 
\begin{eqnarray} 
Y^u_{ij} & = & \frac{m_i^u}{v} \left(
\frac{\cos \al}{\cos \beta} \delta_{ij}+ 
\frac{2}{\sin 2 \beta} (g^u_R)_{ij} \sin (\al-\beta) \right),
\\
 Y^{u-}_{ij} & = & \sum_l (V)^*_{li} ~\frac{\sqrt{2} m_l^u}{v}
 \left( \tan \beta \delta_{lj} 
 - \frac{2}{\sin 2 \beta} (g^u_R)_{lj} \right),
 \\
Y^{ua}_{ij} & = & \frac{m_i^u}{v} \left(  \tan \beta \delta_{ij} 
 -  \frac{2}{\sin 2 \beta}  (g^u_R)_{ij} \right), 
\end{eqnarray}
where $V$ is the Cabibbo-Kobayashi-Maskawa matrix, 
$\al$ is the mixing angle of the neutral Higgs 
fields, and $(g^u_R)^*_{ij}=(g^u_R)_{ji}= (R^{u})^{*}_{i3} (R^{u})_{j3}$.   
Note that the down-type does not have off-diagonal elements like the up-type 
quarks, since we assumed that the down-type quarks carry null or universal 
$U(1)^{'}$ charges. 
The down-sector of the charged, $Y^{d+}_{ij} \ov{\Hat{U}_{Li}}h_0^+ \Hat{D}_{Rj}$, 
is given by $V_{ij} \sqrt{2} m_j^d \tan \beta/v$, and $Y^{da}_{ij}$ 
does not have off-diagonal elements.   

Finally, for the two-Higgs doublet case, one can easily show that 
$|(g_R)_{ut} |^2=(g_R)_{uu}(g_R)_{tt}$ is satisfied.    
In the following, we will treat with $(g_R)_{ij}$ as free parameters  
and concentrate on $(g_R)_{ut}$, $Y_{tu}$, and $Y^{a}_{tu}$ by assuming 
that $D^0$-$\ov{D}^0$ mixing is well suppressed by appropriate parameters. 

\begin{figure}[!t]
\begin{center}
{\epsfig{figure=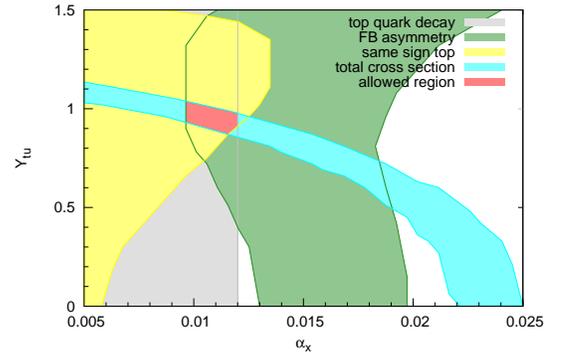,width=0.4\textwidth}}
\end{center}
\vspace{-1.0cm}
\caption{
The favored region for $\alpha_x$ and $Y_{tu}$.
}
\label{fig:axYtu}
\end{figure}

Now we discuss the top FB asymmetry, 
the CDF $Wjj$ excess, and related issues in the framework of our model.
As discussed previously,  we assume that
the $Z^\prime$ boson has only one large off-diagonal element ${g_R}_{ut}$
and two large diagonal elements ${g_R}_{tt},{g_R}_{uu}$ with
a parametrization $\alpha_x = (g^\prime {g_R}_{ut})^2/(4\pi)$.
We assume that ${g_R}_{tt} \gg {g_R}_{uu}$ in order to avoid the strong bound
from the dijet production in the UA2 experiment.
In the Yukawa sector, $Y_{tu}$ and $Y_{tu}^a$ could be $\sim O(1)$
because of the heavy top mass.  

If a new particle has a flavor-changing coupling to the top quark
and is lighter than the top quark, the branching fraction of
the top quark decay to $Wb$
might be significantly altered. To prevent this harmful situation, we 
assume the $Z^\prime$ boson mass $m_{Z^\prime}= 145$ GeV,
the lightest neutral Higgs boson $m_{h}=180$ GeV,
and the lightest pseudo-scalar Higgs boson $m_{a}=300$ GeV.
The mass of the neutral scalar Higgs boson looks like conflict with the recent
CMS bounds in the mass region 149-206 and 300-440 GeV at 95 \% C.L. 
in the SM~\cite{CMS-PAS-HIG-11-011}.
However the bounds would be weaker in our model because new decay channels
of the Higgs boson, for example, $h\to t\bar{u}$, $h\to$ dark matters,
etc., are open. Actually, the branching fraction of the Higgs boson
to $WW$ could be less than $0.5$ in a reasonable parameter space, 
which is enough to evade the current search limits at the LHC.
In Fig.~\ref{fig:axYtu}, the gray  region represents the region in which
the branching fraction of the top quark to $Z^\prime u$ is less than 5 \%.

In our model, the $Z^\prime$ boson contributes to the $t\bar{t}$ production
through its $t$-channel or $s$-channel exchange in the $u\bar{u}\to t\bar{t}$
process, while the scalar and pseudo-scalar Higgs boson are mediated
only by the $t$-channel exchange. For simplicity, we assume $Y_{tu}^a=1.1$.
For numerical analysis, we use CTEQ6m for a parton distribution function
and take the renormalization and factorization scales to be the top quark mass
$m_t = 173$ GeV. The $K$ factor is taken to be $1.3$.
Up to now, the experiments at the Tevatron present the results for 
the $t\bar{t}$ pair production cross-section 
with the smallest errors, $\sigma(t\bar{t}) = ( 7.5 \pm 0.48 )$ pb, 
at the  center-of-momentum energy $\sqrt{s}=1.96$ TeV 
at the Tevatron~\cite{cdf-note-9913}.
In Fig.~\ref{fig:axYtu}, the cyan region corresponds to the allowed region
for the couplings $\alpha_x$ and $Y_{tu}$ in the 1$\sigma$ level.
The $t\bar{t}$ invariant mass distribution in our model follows
the typical pattern from new $t$-channel physics \cite{Jung:2009jz},
but looks closer to the SM prediction 
compared to the $Z^\prime$ only case \cite{Ko:2011di}.

Now we consider the top quark FB asymmetry $A_{\textrm{FB}}$
at the Tevatron. The CDF Collaboration reported 
$A_{\textrm{FB}}^{\textrm{lepton+jets}}= (0.158 \pm 0.075)$
in the lepton+jets channel with an integrated luminosity 
of $5.3$ fb$^{-1}$~\cite{Aaltonen:2011kc}.
A similar deviation from the SM prediction in the dilepton channel
$A_{\textrm{FB}}^{\textrm{dilepton}}=0.42 \pm 0.17$
was reported at CDF~\cite{cdfdilepton}.
Very recently the D0 Collaboration reported 
$A_{\textrm{FB}}=0.196\pm 0.065$ in the lepton+jets channel
with an integrated luminosity of $5.4$ fb$^{-1}$~\cite{Abazov:2011rq}.
All measurements show about 2 or 3 $\sigma$ away from the SM.
For illustration, we use the result in the lepton+jets channel at CDF.
In Fig. \ref{fig:axYtu}, the green region corresponds to the favored region 
for $\alpha_x$ and $Y_{tu}$ in the 1$\sigma$ level.
Here we ignore the  $\sim 5 \%$ asymmetry from 
the next-to-leading order (NLO) contribution
in the SM.  Adding this to our predictions will make the $A_{\rm FB}$ larger.
$A_\textrm{FB}^\textrm{New}$ in the allowed region in Fig.~\ref{fig:axYtu}
is between $0.084$ and $0.12$ without the contribution from the SM NLO.

Models with a light $Z^\prime$ boson or a light scalar boson are 
strongly constrained by the same sign top pair production at 
the LHC ~\cite{Chatrchyan:2011dk}. 
Up to the present, the most stringent bound for the same sign top quark
pair production is given by the CMS Collaboration, 
$\sigma(tt) < 17$ pb~\cite{Chatrchyan:2011dk}, which 
excludes the $Z^{'}$ solution for the top FB asymmetry by Jung et al..  
In our model, the situation becomes completely different, since $h$ and 
$a$ bosons can contribute to  the same sign top pair production though 
their $t$-channel exchanges, in addition to the usual $Z^{'}$ contribution, 
and can reduce the rate for the same sign top pair production due to
the destructive interference among three contributions from $Z^{'}$, 
$h$, and $a$. 
In Fig. \ref{fig:axYtu}, the yellow region shows the allowed region
from this constraint, where the light $Z^{'}$ scenario survives the same
sign top pair constraint due to the destructive interference from $h$ and 
$a$ contributions in the $t$-channel.

In principle, the large flavor changing neutral current in the top sector 
gives rise to a large single top quark production at hadron colliders.
The D0 Collaboration has measured the single top quark production
with $\sigma(p\bar{p}\to t b q+X)=2.90\pm 0.59$ pb 
\cite{Abazov:2011rz}.
However, our model would not be constrained by the measurement.
This is because the main production channels for the singlet top quark are
the $g u\to t Z^\prime$ or $t h$ processes, but 
the branching fractions of $Z^\prime$ and $h$ decays to the $b q + X$ 
state are quite small. Eventually our model would be strongly constrained
if the cross-section in the $p \bar{p} (p) \to t + X$ channel is measured.

In Fig.~\ref{fig:axYtu}, the red region is consistent with all 
the experimental results from the Tevatron and LHC up to now. 
A lot of the parameter regions for $\alpha_x$ and $Y_{tu}$ are excluded,
but there is a region satisfying the strong constrains from the Tevatron
and LHC. If one applies a different set for $m_{Z^\prime}$, $m_h$, etc.,
the allowed region will be slightly altered. However, Fig.~\ref{fig:axYtu}
implies that a model with light $Z^\prime$ only would be
excluded especially from the same sign top pair production result
at CMS.

\begin{figure}[!t]
\begin{center}
{\epsfig{figure=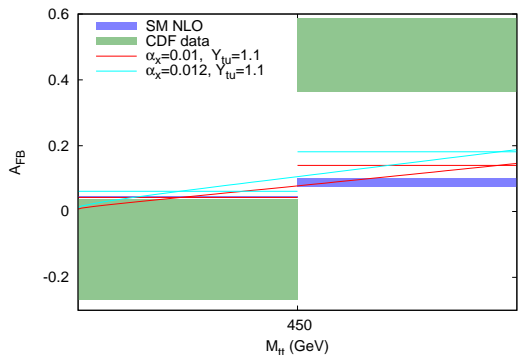,width=0.4\textwidth}}
\end{center}
\vspace{-1.0cm}
\caption{
The top forward-backward asymmetry as a function of the $t\bar{t}$ 
invariant mass. 
}
\label{fig:asym}
\end{figure}

One of the striking features of the CDF data is that
$A_{\textrm{FB}}$ in the $t\bar{t}$ invariant mass region larger than 450 GeV
is over by about $3.4$ $\sigma$ from the SM prediction.
In Fig.~\ref{fig:asym}, we depict our results for the invariant mass
distribution of $A_{\textrm{FB}}$ for two reference parameter sets:
$\alpha_x=0.01, Y_{tu}=1$ (red line or red bin), and
$\alpha_x=0.012, Y_{tu}=1$ (cyan line or cyan bin).
The green and blue bands correspond to the CDF data in the lepton+jets 
channel and the SM prediction from {\sc mc@nlo}, respectively.
Our prediction in the large $t\bar{t}$ invariant mass region is rather smaller
that the CDF data. However, if we include the NLO corrections in the SM,
the prediction would be consistent with the data within about $2$ $\sigma$.
We note that the data in the dilepton channel at CDF imply smaller
$A_{\textrm{FB}}$ in the large invariant mass region, which are consistent with
our prediction.

Finally, we consider the dijet production associated with a $W$ boson
at CDF.  In our model, the most important one is the parton process
$u\bar{b}(b\bar{u}) \to h^\pm \to W^\pm Z^\prime$ with a subsequent decay 
$Z^\prime\to j j$,  where $h^\pm$ is the lightest charged Higgs boson.
The charged Higgs boson has a similar coupling structure to the neutral coupling, 
so that only the $u_R$-$b_L$-$h^+$ and  $b_R$-$u_L$-$h^-$ vertices 
can be as large as that for the $u_R$-$t_L$-$h$. 
The interaction lagrangian for the charged Higgs boson with the $W$ and 
$Z^\prime$ bosons is given by \cite{Ko:2011di}
\begin{equation}
\mathcal{L} = - g^\prime m_W \sin 2\beta H^+ {W^-}^\mu Z^\prime_\mu 
+ h.c. .
\end{equation}
For  $m_{h^\pm}=270$ GeV, we get $\sigma (W Z^\prime) \sim 10$ pb 
$\times \sin^2 2 \beta \lesssim 10$ pb at the Tevatron. 
It would be about $4.5$ pb for $\sin 2 \beta =0.7$ in the range of the CDF 
report, but it could be substantially smaller if  $\sin 2\beta$ becomes smaller.


Before closing, we discuss the anomaly cancellation and possible 
cold dark matter (CDM) for completeness.  
There are a number of solutions for anomaly cancellation, 
and a simple way is to add one extra generation with the opposite chirality 
and the same charge assignment as the third generation 
in the $(u_1,u_2,u_3)=(0,0,1)$ case.
This extra generation allows the mass mixing between the SM fermions 
and extra fermions, such as $\ov{D'_L}D_{Ri}$, 
where $D_L^\prime$ is the extra left-handed down-type quarks,
so that we assume that such unfavored terms are small enough to 
avoid large flavor-changing neutral-current contributions. 
If we consider the case $(u_1,~u_2, ~u_3)=(-1,~0,~1)$, we need $3$ Higgs field 
whose charges are $(-1,~0,~1)$, but  we do not need such extra family. 
Instead, we need extra chiral fermions, $(q_{LI},~q_{RI})$ where $I=1,2$, 
to cancel the $U(1)_YU(1)'^2$ anomaly. 
When they are $SU(3)_c$ fundamental representations 
with $U(1)_Y$ charges, $1/3$,
the $U(1)'$ charges of $(q_{LI},~q_{RI})$ must be defined 
as $(3/2,~1/2)$ and $(-3/2,~-1/2)$,
and they also require a extra scalar, which could 
be a good CDM candidate \cite{Ko:2011di}.
  

In this article, we presented  $U(1)'$ flavor models with flavor dependent
$Z^{'}$ couplings only to the RH up-type quarks. 
The $U(1)'$ charges of the SM quarks and Higgs fields are chosen in order to 
generate the renormalizable Yukawa couplings for the up-type quarks. 
We could explain $\sigma_{t\bar{t}}$, the top FB asymmetry,  
the CDF $Wjj$ excess, etc., without any conflict with other phenomenological 
constraints with $m_{Z^{'}} \sim 145$ GeV, $m_h \sim 180$ GeV, and 
$m_{H^+} \sim 270$ GeV.
In particular, the new contributions from $h$ and $a$ in the $t$-channel could
help the models evade the stringent constraint from the same sign top 
pair production. 
This aspect is new to our flavor dependent $U(1)^{'}$ model and would be 
generic to other models with new chiral gauge interactions. 
There always appear new scalar bosons with flavor dependent couplings for 
renormalizable Yukawa couplings, and  their effects on the top FB asymmetry 
must be included for theoretical consistency. It is not enough to consider the 
$Z^{'}$ contribution to the top FB asymmetry and the same sign top pair productions.
On the other hand, CDM phenomenology from the new matter contents to achieve 
anomaly cancellation is more model-dependent, and the details will 
depend on each specific case. 
For our particular choice of $U(1)'$ charges shown in Eq.~(\ref{assign}), new 
$Z^{'}$ couples only to the RH up-type quarks and not to the  down-type quarks.  
Therefore the decay $Z^{'} \rightarrow b\bar{b}$ is absent in our model and 
the CDF $Wjj$ within our model will disappear if the $b$-tagging is applied to the dijet. 
However this could change if we assign flavor universal $U(1)'$ couplings to 
the down-type quarks. 

\begin{acknowledgments}
We are grateful to Sunghoon Jung for useful discussions. 
The work of PK is supported in part by the SRC program of 
the National Research  Foundation, through KNRC at Seoul National University. 
The work of CY is supported by Basic Science Research Program
through NRF (2011-0022996).
\end{acknowledgments}

\end{document}